\documentclass[twocolumn,showpacs]{revtex4-1}
\usepackage{color}
\usepackage{graphicx}
\usepackage{dcolumn}
\usepackage{amsmath}

\begin{document}
\bibliographystyle{revtex}

%\preprint{HEP/123-qed}

\title[Short Title]{Cold nuclear matter}

\author{C.O. Dorso, P.A. Gim\'enez Molinelli and J.I. Nichols}

\affiliation{Departamento de F\'isica, FCEN, Universidad de Buenos
Aires, N\'u\~nez, Argentina}
\affiliation{IFIBA-CONICET}

\author{J.A. L\'opez}

\affiliation{Department of Physics, University of Texas at El
Paso, El Paso, Texas 79968, U.S.A.}

\date{\today}

\begin{abstract}

The behavior of nuclear matter is studied at low densities and
temperatures using classical molecular dynamics with three
different sets of potentials with different compressibility.
Nuclear matter is found to arrange in crystalline structures
around the saturation density and in non-homogeneous (i.e.
pasta-like) structures at lower densities.
Similar results were obtained with a simple Lennard-Jones 
potential.
Finite size effects are analysed and the existence of the 
non-homogeneous structures is shown to be inherent to the use of 
periodic boundary conditions and the finitude of the system.
For large enough systems the non-homogeneous structures are 
limited to one sphere, one rod or one slab per simulation cell, 
which are shown to be minimal surface structures under cubic 
periodic boundary conditions at the corresponding volume fraction.
The relevance of these findings to the simulations of neutron star 
and supernovae matter is discussed.

%   Using
% topological tools the non-homogeneous shapes were found to evolve
% into the gas-liquid phase as a function of the temperature.  The
% observation of the transition from the nuclear pasta to the
% liquid-gas mix is considered to be a valuable addition to the
% phase diagram of nuclear matter.

\end{abstract}

\maketitle

\section{Introduction}\label{intro}

The study of cold nuclear matter at subsaturation densities is
important for a variety of topics which include the equation of
state of nuclear matter, phenomena related to the study of heavy
ion reactions, the structure of neutron star crusts, etc.
Pioneering studies based on the
compressible liquid drop model~\cite{lamb,Lattimer}, Hartree-Fock
method~\cite{bonche}, and energy minimization
techniques~\cite{raven,koonin} showed that transitions from
spherical nuclei to shapes such as rods, slabs, tubes, spherical
bubbles, are to be expected in cold nuclear matter.  More recent
investigations~\cite{raven,21,koonin,23,24,25,26,27,28,29,30,dor12,
dor12A}
have used dynamical methods to study such transitions in neutron
star crust environments.

% A major component in all of these studies has been the presence of
% an all embedding electron gas which, in neutron stars, is expected
%to be produced by an abundance of  $\beta$-decays.
% In fact, a
Coulomb interaction has been thought of as an essential ingredient
for the formation of the rich ``pasta'' like structures.
Koonin~\cite{koonin}, for instance, explained the transitions
between different topologies in terms of a competition between a
short-range nuclear surface energy which becomes minimized through
aggregation, and a long-range Coulomb energy which gets reduced by
an opposite dispersion.  More recent studies of Horowitz and
coworkers~\cite{horo_lambda} re-examine this assertion as an
example of {\it frustration}, a phenomenon that emerges from the
impossibility to simultaneously minimize all interactions, and
yields a large number of low-energy configurations. Indeed in all
of the previous studies listed, the Coulomb interactions have been
approximated either by electrically grounded
surfaces~\cite{raven}, uniform electric charge
densities~\cite{koonin}, Thomas-Fermi screened Coulomb potential
(see, e.g.~\cite{7}) or by an Ewald summation~\cite{wata-2003}.

The importance the Coulomb repulsion has on the formation of the
pasta-like structures prompted a previous study~\cite{dor12-2}
which dissected the role of this interaction as a function of
density, temperature and isotopic content through the use of
classical molecular dynamics simulations at fixed volume and
number of particles.  In an unexpected outcome, however, such
study, which varied the strength of the electric interaction from
full to none, observed that the pasta structures, namely,
``gnocchi'', ``spaghetti'', ``lasagna'' and their anti-structures
(i.e. those obtained by replacing particles by holes and
viceversa), existed even in the absence of Coulomb interaction.
Although the structures appeared modified somewhat in their
overall scales, topology and location in the density-temperature
plane, the nuclear potential seemed to be sufficient to give rise
to a rich pasta-like structure in nuclear systems at subsaturation
densities and low temperatures under periodic boundary conditions.
Nuclear matter at subsaturation densities and high temperatures
decomposes into a mix of liquid and gaseous phases~\cite{Lop00},
at lower temperatures ($T\lesssim 1 \ MeV$), however, it seems to
self-assemble into pasta-like objects even without Coulomb
interaction. This gives rise to very interesting questions that
are addressed in the present work.

The approach to be followed is a combination of classical
molecular dynamics of large (but finite) systems under periodic
boundary conditions, topological analysis tools to study the
structure of infinite nuclear matter at subsaturation densities
and very cold temperatures, and schematic geometrical
considerations. In the next section the classical molecular
dynamics model used will be briefly reviewed for completeness.
Section~\ref{pandha-med} presents a detailed study of the
structure of symmetric nuclear matter with medium compressibility,
followed in Section~\ref{other} by similar studies with other
potentials that have been used in the past to study nuclear
systems.  The role of finite size effects is briefly discussed in
Section~\ref{origin}, and a summary of the main results is
presented in Section~\ref{concluding} along with some concluding
remarks. Visual representations of simulated systems where made
using VMD\cite{cite_vmd}.

\section{Classical molecular dynamics}\label{cmd}
This work uses a classical molecular dynamics ($CMD$) model to
study infinite nuclear matter at low temperatures and
subsaturation densities; the use of molecular dynamics to study
nuclear reactions was pioneered by Wilets and
coworkers~\cite{wilets} and advanced by
Pandharipande~\cite{pandha} and others~\cite{lop-lub,dor-ran}.
Recently, classical molecular dynamics models have been used to
study cold nuclear matter in neutron star crusts
environments~\cite{horo_lambda,P14,P15,P2012}; in particular the
$CMD$ model, which was developed to study nuclear
reactions~\cite{14a,Che02,16a,Bar07,CritExp-1,CritExp-2,TCalCur,
EntropyCalCur,8a,Dor11}, has been adapted to study infinite
nuclear systems under such conditions~\cite{dor12,dor12A,dor12-2}.

In this study, the trajectories of the nucleons are governed by
classical equations of motion dictated by forces produced by the
Pandharipande~\cite{pandha} potentials:
\begin{align*}
V_{np}(r) &= V_{r}\left[ exp(-\mu _{r}r)/{r}-exp(
-\mu_{r}r_{c})/{r_{c}}\right] - \\
&-V_{a}\left[ exp(-\mu _{a}r)/{r}-exp(-\mu
_{a}r_{c})/{r_{c}}
\right] \\
V_{NN}(r)&=V_{0}\left[ exp(-\mu _{0}r)/{r}-exp(-\mu _{0}r_{c})/{
r_{c}}\right] \ , \label{2BP}
\end{align*}
where the attractive potential between a neutron and a proton is
$V_{np}$, and  the repulsive interaction between similar nucleons
($nn$ or $pp$) is $V_{NN}$; they both use a cutoff radius of
$r_c=5.4$ $fm$ after which the potentials are set to zero. The
Yukawa parameters $\mu_r$, $\mu_a$ and $\mu_0$ were
phenomenologically adjusted by Pandharipande to yield a saturation
density of $\rho_0=0.16 \ fm^{-3}$, a binding energy
$E(\rho_0)=16$ MeV/nucleon and a compressibility (actually, bulk
modulus) listed in~\cite{pandha} as $250 \ MeV$ for the ``Medium''
model, and $535 \ MeV$ for the ``Stiff'''~\cite{pandha}.

The trajectories of all nucleons are obtained by solving the
classical equations of motion using a symplectic Verlet algorithm
with energy conservation of $\mathcal{O}$($0.01\%)$. To mimic an
infinite system $A=1728$ to $A=13824$ nucleons were placed in
cubic cells under periodic boundary conditions. We focus on
isospin symmetric systems $x=z/A=0.5$. The number densities were
enforced by placing a fixed number of nucleons in cubical boxes
with sizes selected to adjust the density. To study systems at
different temperatures, the nuclear matter is force-heated or
cooled using the Andersen thermostat~\cite{andersen} to control
the temperature. Previous studies already presented samples of
structures obtained through this method~\cite{dor12,dor12A}. In
this study of cold nuclear matter the range of densities is
selected to be $0.01 \ fm^{-3} \le \rho \le 1.25\rho_0$, and that
of densities is $0.001\lesssim T \lesssim 1.0 \ MeV$.

\begin{figure}  %figure 1
\begin{center}
\includegraphics[width=3.4in]{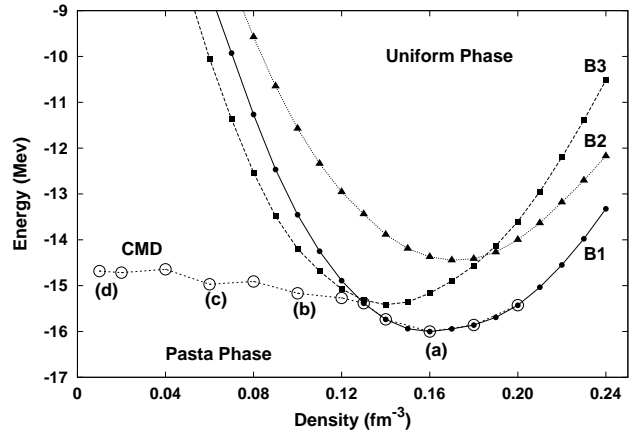}
\end{center}
\caption{Binding energy per nucleon for systems obtained with the
Pandharipande medium potential with simple cubic ($B1$), body
centered cubic ($B2$) and diamond ($B3$) crystal lattices, and
using molecular dynamics ($CMD$) at $T=0.001 \ MeV$ ($CMD$).  The
structures corresponding to the four labelled points (``A''
through ``D'') are shown in Figure~\ref{fig2}. Notice that around
saturation density the $CMD$ results agree with those of the
simple cubic lattice.} \label{fig1}
\end{figure}

\begin{figure}  %figure 2
\begin{center}
\includegraphics[width=3.2in]{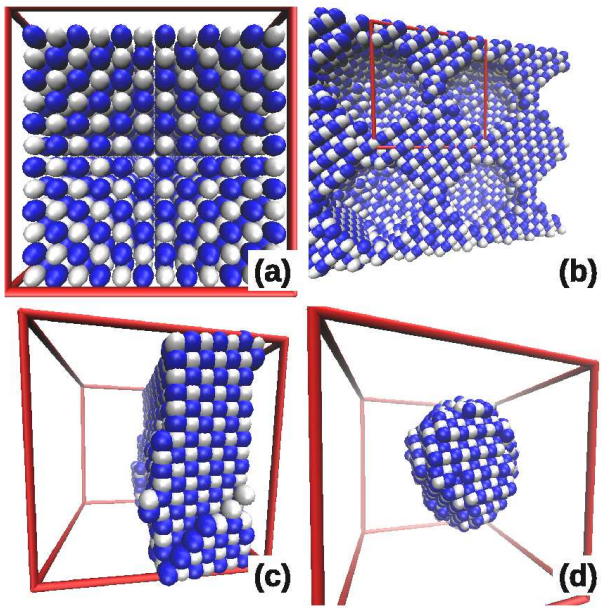}
\end{center}
\caption{(Color online) Structures corresponding to the labelled
points of Figure~\ref{fig1}. Point A corresponds to a formation in
the regular ($B1$) lattice, while the rest of the points are
non-homogeneous structures.}\label{fig2}
\end{figure}

The procedure we follow is twofold. To study the uniform phase
exactly at zero temperature, a crystalline structure at a given
density is constructed and its energy per nucleon calculated by
direct summation between all nucleons. The dependence of the
binding energy on the density is explored by changing the lattice
parameter different, the energy versus density curve is shown in
Fig.~\ref{fig1}. This procedure, identical to the one used by
Pandharipande~\cite{pandha}, produces the characteristic ``U'',
with a minimum signalling the saturation or normal nuclear
density.

\begin{figure}  %figure 5
\begin{center}
\includegraphics[width=3.4in]{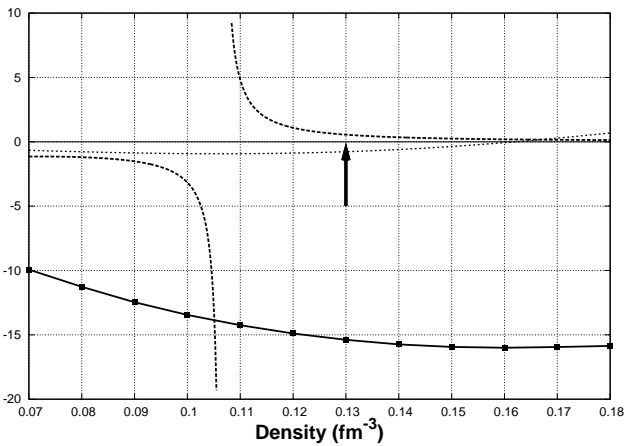}
\end{center}
\caption{Energy (full lines with squares), pressure (dashed line)
and compressibility (dotted line) for the uniform $B1$ lattice of
the Medium potential. The arrow points at the density at which the
system departs from homogeneity.} \label{fig_pres_med}
\end{figure}

Pandharipande \textit{et al.} assumed that the nucleons in the
ground state were arranged as a simple cubic lattice for the whole
range of densities, but that is not necessarily the case as we
shall see in Section~\ref{other}. Given that nuclear matter is
composed of two kinds of particles (neutrons and protons), the
crystal geometries they adopt are similar to those formed by
binary alloys. As we will see, the relevant crystal structures for
the present case are the $B1$ (a simple cubic lattice in which
every first neighbor of a proton is a neutron and viceversa, used
by Pandharipande \textit{et al}), the $B2$ (a $BCC$ lattice), and
the $B3$ (a diamond lattice with nucleons arranged so that every
first neighbor of a protons is a neutron and viceversa).

\begin{figure}[b]  %figure 3
\begin{center}
\includegraphics[width=3.4in]{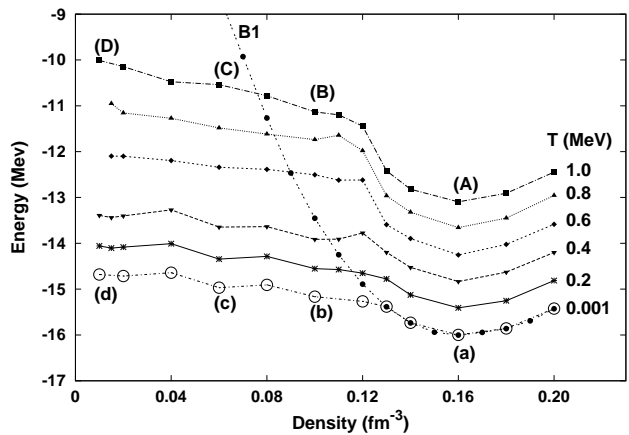}
\end{center}
\caption{Binding energy per nucleon for systems obtained with the
Pandharipande medium potential at the listed temperatures.}
\label{fig3}
\end{figure}

The second method uses $CMD$ starting from a random positioning of
a fixed number of nucleons in a central cubic cell under periodic
boundary
conditions, with a Maxwell-Boltzmann velocity distribution
corresponding
to a given initial temperature.  The system is then equilibrated at
a high temperature ($T\gtrsim 2 \ MeV$) and then brought down to
the final desired temperature using the Andersen thermostat
procedure in small temperature steps.  $T=0.001 \ MeV$ is taken as
zero temperature for the $CMD$ calculations, other values explored
are in the range $0.001 \ Me V \le T \le 1 \  MeV$. After
reaching equilibrium, the analysis tools described in~\cite{dor12A}
are used to visualize and characterize the produced structures.

It is with these tools that cold nuclear matter will be studied under
a variety of conditions and for several potentials.

% Figure~\ref{fig1} energy curves for the Pandharipande medium
% potential
% as produced by $CMD$ and for each uniform crystal structure as a
% function of density.

\begin{figure}  %figure 4
\begin{center}
\includegraphics[width=3.2in]{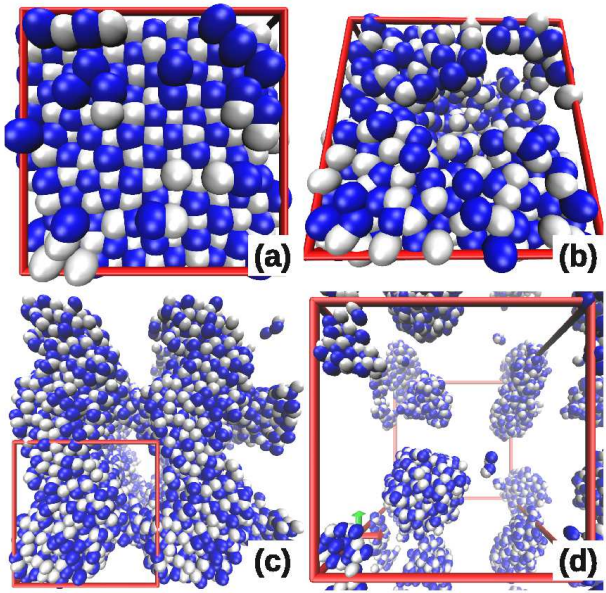}
\end{center}
\caption{(Color online) Structures corresponding to the labelled
points of Figure~\ref{fig3} obtained with the Pandharipande medium
potential at $T=1.0 \ MeV$.}\label{figT1}
\end{figure}

\section{Medium compressibility matter}\label{pandha-med}

Figure~\ref{fig1} shows the near zero temperature results for
symmetric ($x=0.5$) matter interacting through the Pandharipande
medium potential (see figure caption for details).  As can be
seen, $CMD$ reproduces the simple cubic ($B1$) lattice
calculations up to a density of $\rho\approx 0.13 fm^{-3}$, while
for lower densities the systems breaks into pasta-like objects of
different shapes. Also shown in Figure~\ref{fig1} are the energies
expected for a uniform body centered cubic ($B2$) and diamond
($B3$) lattice structures which, being higher in energy, do not
correspond to the $T=0$ case. Figure~\ref{fig2} shows the
structures corresponding to the four densities labelled from ``A''
to ``D'' in Figure~\ref{fig1}.

The structures of Figure~\ref{fig2} resemble those obtained by
Williams and Koonin in 1985 with a static mean field
model~\cite{koonin}. Placing nuclear matter in a periodic simple
cubic lattice at a given number density of nucleons, the system
was allowed to relax freely until a local energy minimum was
achieved. Although the method did not take into account the
possibility of having free nucleons and used a fixed geometry by
construction, it yielded results comparable to those obtained by
the dynamical method used in this study.  For instance, structure
``C'' in Figure~\ref{fig2} --which corresponds to what
in~\cite{dor12A} was dubbed as ``lasagna''-- was also found by
Koonin~\cite{koonin} (``alternating slabs of matter and vacuum'')
and Ravenhall~\cite{raven}. A significant difference, however, is
that both Ravenhall and Koonin included the Coulomb interaction as
a main ingredient, which we do not do; this will be addressed in
Section~\ref{origin}

As an aside, the $B1$ curve of Figure~\ref{fig1} can be used to
perform a polynomial fit around its minimum to extract an analytic
expression, which can then be used to calculate the
compressibility at around the saturation density through
$K=9\rho^2_0 [d^2(E/A)/d\rho^2]_{\rho_0}$.  In the case of the
medium Pandharipande potential the value of the bulk modulus is
found to be $283 \ MeV$, comparable with the value of $250 \ MeV$
quoted by its creators.

In Figure~\ref{fig_pres_med}, we plot the energy, pressure and
compressibility for the homogeneous $B1$ lattice of
Pandharipande's Medium potential. The pasta-like structures are
found in a mechanically unstable (negative pressure) density
region, but well above the divergence
in compressibility.% that signals the liquid-vapor instability.

\begin{figure}  %figure 6
\begin{center}
\includegraphics[width=3.5in]{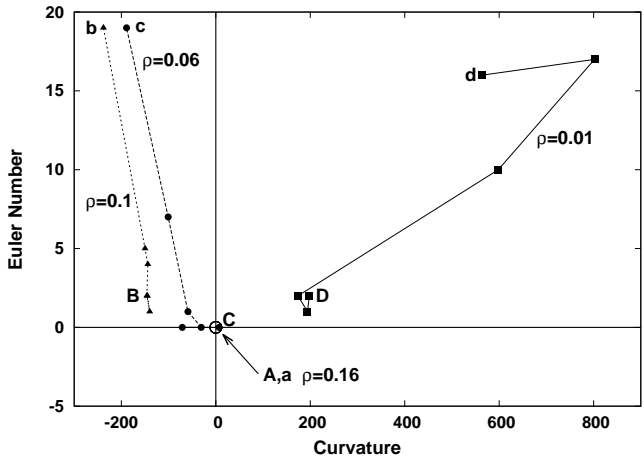}
\end{center}
\caption{Curvature - Euler coordinates of the structures of
Figure~\ref{fig3}.  The lines connect points with the same
densities but temperatures varying from $T=0.001 \ MeV$ to $1.0 \
MeV$.} \label{figCE}
\end{figure}

%It was a case of good judgement that
% Williams and Koonin chose a simple cubic as their basis, have they
% preferred to use $FCC$ or $BCC$ instead, their saturation density
% and binding energy would have been off from the correct values.
% Likewise, their assumption of no free neutrons turned out to be
% correct for the case of $x=0.5$.

\begin{table}
\centering  % used for centering table
\caption{Classification Curvature - Euler}
\begin{tabular}{c|| c | c | c} % centered columns (4 columns)
\hline                  % inserts single horizontal line
& Curvature $<0$ & Curvature $\sim 0$ & Curvature $>0$ \\ % inserting
% body of the table
\hline \hline                        %inserts double horizontal lines
Euler $>0$ & Anti-Gnocchi &  & Gnocchi  \\
Euler $\sim0$ & Anti-Spaghetti & Lasagna & Spaghetti  \\
Euler $<0$ & Anti-Jungle Gym &  & Jungle Gym  \\ [1ex]      % [1ex]
% adds vertical space
\hline %inserts single line
\end{tabular}
\label{table1} % is used to refer this table in the text
\end{table}

Continuing with the study at higher temperatures,
Figure~\ref{fig3} shows the same type of results as
Figure~\ref{fig1} for temperatures $0.001 \ MeV \le T \le 1.0 \
MeV$.  As it can be seen in the figure, at densities $\rho \gtrsim
0.13 \ fm^{-3}$ and at all of these temperatures the curves follow
the ``U'' shape characteristic of the uniform $T=0$ crystalline
phase.  As in the case of zero temperature, if the density
decreases below, say $\rho \lesssim 0.13 \ fm^{-3}$ the systems
again move away from the uniform phase forming complex
arrangements. Figure~\ref{figT1} shows the structures obtained for
the highest temperature studied ($T=1.0 \ MeV$) for the four
densities labelled from ``a'' to ``d'' in Figure~\ref{fig3}.

These non-homogeneous structures being formed in the low density
region can be characterized using the mean curvature and Euler
number.  As explained in detail in~\cite{dor12A}, different
structures have distinct values of these variables and, in
general, follow the pattern outlined in Table~\ref{table1}; as a
reference, the perfect crystals formed at $T=0$ and $\rho \gtrsim
0.13 \ fm^{-3}$ ({\it v.g.} point ``A'' in Figures~\ref{fig1}
and~\ref{fig3}) are formally uniform and infinite because of the
periodic boundary conditions imposed, hence they have no surfaces
and null Euler characteristic.

The curvature-Euler coordinates of the labelled structures of
Figure~\ref{fig3} are presented in Figure~\ref{figCE}; points
joined by lines all have the same densities but their temperatures
vary through $T=0.001, \ 0.1, \ 0.5, \ 0.6, \ 0.8$ and $1.0 \ MeV$
with the letter labels corresponding to those of
Figure~\ref{fig3}. Again, the cases ``A-a'' at normal density
($\rho_0$) correspond to uniform crystalline structures so all
have zero curvature and Euler number at the temperatures studied.
In the case ``B-b'' of density $0.1 \ fm^{-3}$ the almost
spherical bubbles at $T=0.001 \ MeV$ become distorted at higher
temperatures. Perhaps the most interesting case is ``C-c'' at
$\rho=0.06 \ fm^{-3}$, which goes from being a perfect ``lasagna''
to a ``jungle-gym'', a complex of punctured lasagna joined by
columns. Structures ``D-d'', on the other side, go from spherical
``gnocchi'' to deformed droplets in what can be considered as the
beginning of the transition from the pasta phase to the liquid-gas
mixed phase.

Jumping a little ahead, it is worth mentioning that the previous
results corresponding to non-homogeneous structures are affected
by the use of period boundary conditions in a non-trivial manner.
This will be discussed in more detail in Section~\ref{origin}.

\begin{figure}  %figure 7
\begin{center}
\includegraphics[width=3.4in]{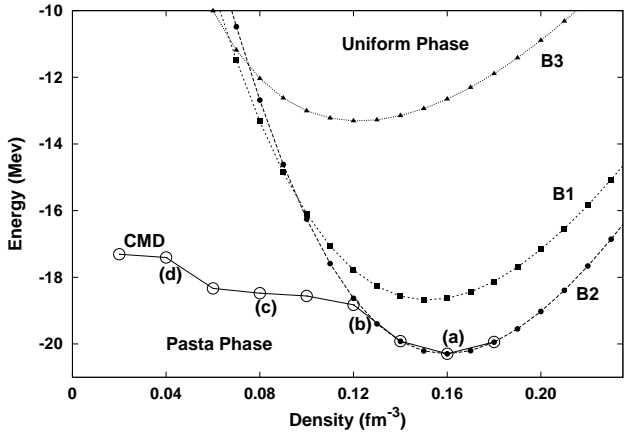}
\end{center}
\caption{Binding energy per nucleon for systems obtained with the
Simple Semiclassical Potentials for crystalline lattices with
$B1$, $B2$ and $B3$ lattice geometries, and using molecular
dynamics at $T=0.001 \ MeV$ ($CMD$).  The structures corresponding
to the four labelled points (``A'' through ``D'') are shown in
Figure~\ref{horostruc}.} \label{horo-ener}
\end{figure}

\section{Higher compressibility potentials}\label{other}

As the comportment presented in the previous section is bound to
be potential-dependent, it is instructive to repeat the study
using other nuclear interactions to extract generalities of the
behavior of nuclear matter at low temperature and subsaturation
densities. In particular, the study uses potentials with higher
values of the compressibility that have been used in the past to
study nuclear matter.

\subsection{A simple semiclassical potential}\label{horo}

A higher compressibility set of potentials that has been used for
a variety of studies of nuclear matter~\cite{P14,P15,P2012} is the
one described by its creators~\cite{horo_lambda} as a ``simple
semiclassical potentials'' ($SSP$).  In summary, the $SSP$ is
composed of:
\begin{eqnarray*}
V_{np}(r) &=& a e^{-r^2/\Lambda} +[b- c] e^{-r^2/2\Lambda} \ , \\
V_{NN}(r)&=& a e^{-r^2/\Lambda} + [b+c] e^{-r^2/2\Lambda} \ ,
\label{horopot}
\end{eqnarray*}
where, again, the potential between a neutron and a proton is
attractive, and that between like particles is repulsive.  The
parameters $a$, $b$, $c$, and $\Lambda$ have been adjusted to have
the proper energy and density scales to mimic nuclear matter.
Notice that, at a difference from the Pandharipande potentials,
the $SSP$ potentials do not have repulsive hard cores.

\begin{figure}  %figure 8
\begin{center}
\includegraphics[width=3.2in]{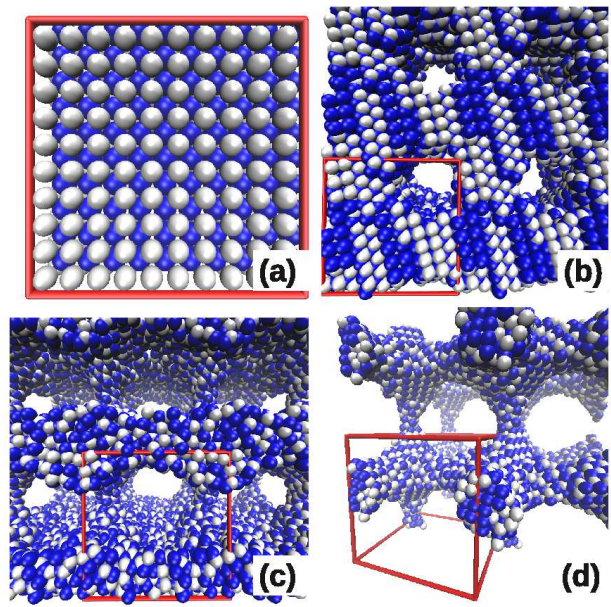}
\end{center}
\caption{(Color online) Structures obtained with the $SSP$
corresponding to the labelled points of Figure~\ref{horo-ener}.
Points A and B correspond to formations in the regular lattice,
while the other points are pasta structures. Panel D resembles the
Schwarz P-Surface, a minimal triply periodic
surface}\label{horostruc}
\end{figure}

Figure~\ref{horo-ener} shows the binding energies per nucleon
obtained from symmetric systems constructed with the $SSP$ in
crystalline lattices with $B1$, $B2$ and $B3$ crystal geometries.
When used in our molecular dynamics code at $T=0.001 \ MeV$ --and
without the Coulomb potential-- the $SSP$ produces structures with
the energies labelled in Figure~\ref{horo-ener} as $CMD$.
Interestingly, at saturation density this potential produces a
$B2$ crystal instead of the $B1$ produced by the lower
compressibility model used in Section~\ref{pandha-med} .

\begin{figure}[t]  %figure 9
\begin{center}
\includegraphics[width=3.5in]{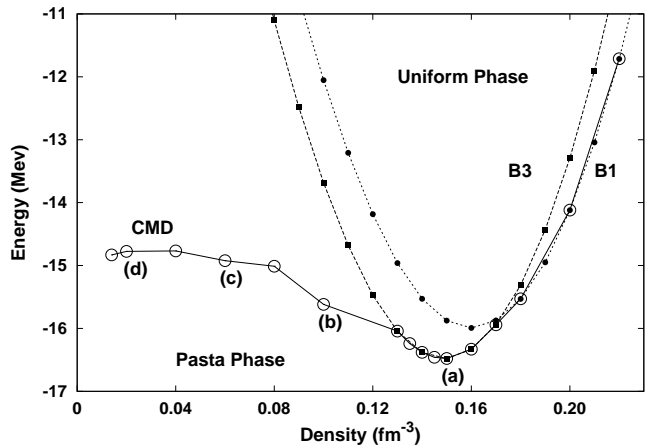}
\end{center}
\caption{Binding energy per nucleon for systems obtained with the
Pandharipande stiff potential for crystalline lattices with
 $B1$ and $B3$ crystal geometries, and using
molecular dynamics at $T=0.001 \ MeV$ ($CMD$).  The structures
corresponding to the four labelled points (``A'' through ``D'')
are shown in Figure~\ref{stiffstruc}.} \label{stiff}
\end{figure}

Noticeably, the ``normal'' density point of this potential is at
the correct value ({\i.e.} at $\rho=0.16 \ fm^{-3}$) but at the
lower binding energy of about $-20.3 \ MeV$.  Again, using a
polynomial fit to the bottom part of the ``U'' of the $B2$ curve
of Figure~\ref{horo-ener}, allows us to extract a value of the
compressibility of the order of $418 \ MeV$, much higher than the
value of the Pandharipande medium potential.

As in the case of the medium Pandharipande potential, at densities
$\rho\gtrsim 0.12 \ fm^{-3}$ the $CMD$ results agree with the
lowest-energy crystalline structure, which in this case is the
$B2$, and at lower densities the systems form pasta-like objects.
The structures corresponding to the four labelled points (``A''
through ``D'') are shown in Figure~\ref{horostruc}.

\begin{figure}  %figure 10
\begin{center}
\includegraphics[width=3.2in]{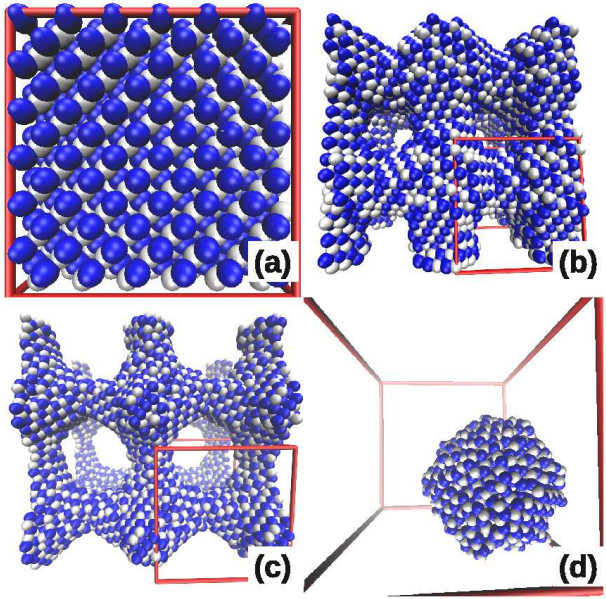}
\end{center}
\caption{(Color online) Structures corresponding to the labelled
points of Figure~\ref{stiff}. Again, point A corresponds to
formations in the regular lattice, while points B through D are
pasta structures.}\label{stiffstruc}
\end{figure}

Except for minor differences, the scenario that the $SSP$ presents
is very similar to that obtained with the Pandharipande medium
potential.  Namely, it is possible to obtain rich pasta-like
structures in an $SSP$ medium at cold temperatures and
subsaturation densities without the modulating effect of the
Coulomb interaction.

\subsection{Stiff Pandharipande potential}\label{stiffsection}

The same calculations are now repeated for the stiff version of
the Pandharipande potential, see~\cite{pandha}. This interaction
produces the $E/A$ versus $\rho$ shown in Figure~\ref{stiff} for
symmetric nuclear matter and zero temperature in $B1$ and $B3$
crystals; the $B2$ structure produced energies higher than the
scale of the figure. Also presented are the binding energies of
the structures obtained with $CMD$ at $T=0.001 \ MeV$.

In this case the saturation point for cold matter occurs at
$\rho=0.15 \ fm^{-3}$, with an energy of $16.5\ MeV$ and for a
$B3$ structure. The authors of the potential did not realize that
the lowest energy geometry was a $B3$ lattice and assumed a $B1$
structure in their estimation of the compressibility which led to
an incorrect value. Using a polynomial fit of the $B3$ curve to
estimate the compressibility at the saturation point, as done with
the other two potentials, yields a value of about $494 \ MeV$,
somewhat smaller than the value of $535 \ MeV$ that Pandharipande
obtained using a $B1$ lattice~\cite{pandha}.

Furthermore, at a density of about $0.17\ fm^{-3}$, this potential
presents a solid-solid phase transition between $B1$ and $B3$
structures. The $CMD$ simulations correctly show that at $\rho
\gtrsim 0.17 \ MeV$ the preferred crystalline structure is the
$B1$, while for lower densities in the range $0.13 \lesssim \rho
\lesssim 0.17 \ fm^{-3}$ it is $B3$. The departure from the
homogeneous phase into the pasta structures starts at densities
smaller than $\rho \approx 0.13 \ fm^{-3}$; the structures
corresponding to the four labelled points (``A'' through ``D'')
are shown in Figure~\ref{stiffstruc}.

Again, except for minute differences, stiff nuclear matter also
produces pasta formations at cold temperatures and subsaturation
densities without any Coulomb interaction.

\section{Pseudo pasta}\label{origin}

Historically, the existence of pasta phases in nuclear matter has
been attributed to the competition between surface and Coulomb
energies. While the short-ranged attractive nuclear interaction
drives the system to a minimum surface configuration, the
long-ranged repulsive Coulomb interaction drives protons as far
away from each other as possible producing the non-homogeneous
pasta structures. However, as we have seen in the previous
sections and in previous studies~\cite{dor12-2}, pasta-like
structures can be found in molecular dynamics simulations with
attractive-repulsive nuclear potentials without Coulomb
interaction.

To elucidate on the origin of these pasta-like structures, this
section will first demonstrate that such structures can be
generated in single-component systems interacting through a single
potential. Since this proves that the attractive-repulsive
interplay of forces cannot be responsible for the formation of
these pasta-like structures, we then proceed to study some
geometrical aspects of the simulations which appear to bear a
large share of the responsibility for the formation of these
structures.

\subsection{Lennard Jones pasta}\label{LJ}
Pasta, as expected to exist in neutron star crusts, originates
from the competition between short-range attractive nuclear
interaction and long-range Coulomb repulsion. Due to such proposed
origin, pasta structures were not expected to exist in pure
nuclear matter but, as seen in the previous sections, they indeed
appear in molecular dynamics simulations. In nuclear matter,
presumably, one could blame the formation of the pasta structures
on the interplay between the attractive and repulsive parts of the
nuclear interactions; since such potentials are of the same range,
that would rule out the phenomenon of frustration (i.e. the
impossibility of obtaining a minimization of all forces at the
same time) as the origin of the pasta. Motivated by this startling
discovery, we decided to simplify the systems treated down to a
one-component one-potential case to see the structures that could
be obtained.

We adapted $CMD$ to perform simulations using a simple
Lennard-Jones ($LJ$) potential under the same conditions of number
density and system sizes as those for the nuclear case. The
interaction potential between any two ``nucleons'' is given by
\begin{displaymath}
 V_{LJ}(r) = 4\epsilon \left[ \left( \frac{\sigma}{r} \right)^{12} - \left(
\frac{\sigma}{r}\right)^6 \right] \ ,
\end{displaymath}
with $\sigma$ and $\epsilon$ chosen to have the minimum energy in
$-16$ in arbitrary units and at $\rho=0.16$ in a close-packed
lattice; conditions similar to those of the Pandharipande medium
potential of Section~\ref{pandha-med}.

\begin{figure}  %figure 16
\begin{center}
\includegraphics[width=3.2in]{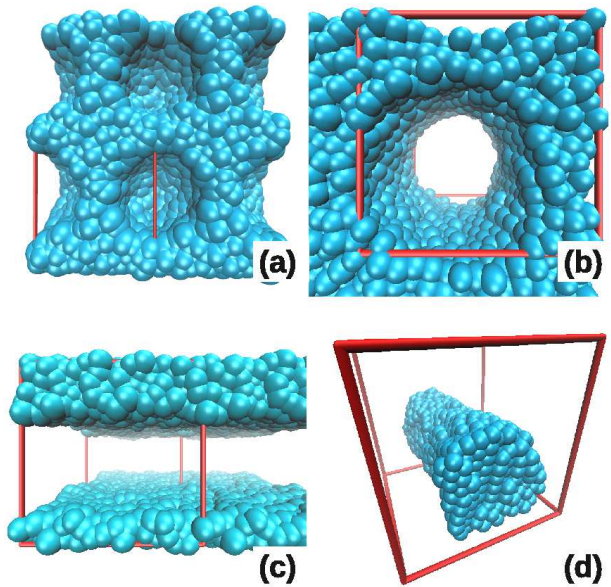}
\end{center}
\caption{(Color online) Structures obtained with a Lennard-Jones
potential}\label{lenardstruc}
\end{figure}

Figure~\ref{lenardstruc} shows some of structures obtained with
the $LJ$ potential, the resemblance to the nuclear matter
structures is striking; as it turns out, the formation of
pasta-like structures had been observed in grand canonical Monte
Carlo simulations of the liquid-vapor coexistence region of $LJ$
systems~\cite{binder2012}. Given that there is no
attractive-repulsive interplay of forces in a one component $LJ$
fluid, these pasta-like structures can only arise from the
geometrical features of the simulation, i.e. from finite size
effects as well as volume and surface considerations.

\subsection{Geometric considerations}

A recurrent question of numerical calculations is the validity of
the results given the finite size of the system treated.  In
uniform systems this concern is readily addressed with the use of
periodic boundary conditions imposed on a cell much larger than
the range of the inter-particle potential. This practice, however,
fixes the maximum size of density fluctuations to be of the size
of the simulation cell and, although this is not a problem for
homogeneous media, simulations of non-homogeneous systems are
affected by this artificial periodicity.

A characteristic worth investigating is that in all of the
pasta-like structures obtained both with pure nuclear matter (cf.
Figures~\ref{fig2},~\ref{figT1},~\ref{horostruc}
and~\ref{stiffstruc}) and with a $LJ$ medium
(Fig.~\ref{lenardstruc}) only one structure appears to be formed
in each cell independent of the cell size. This effect has also
been observed in grand canonical Monte Carlo simulations of $LJ$
pasta-like structures~\cite{binder2012}, which found only one
spherical drop, one rod, one slab, one cylindrical tube or one
spherical bubble per cell, except in densities during transitions.
To examine this intriguing result we look in turn at the volume
and surface of structures obtained at zero temperature with the
medium compressibility potential of Section~\ref{pandha-med}.

In our simulations at zero temperature and with $x=0.5$ the number
of particles is fixed and the cell volume is adjusted to yield the
desired density. As it can be seen in Fig.~\ref{fig1} the binding
energies per nucleon maintains a relatively constant value, which
implies that the local densities of matter bound in pasta-like
structures correspond roughly to the saturation density as
conjectured by Ravenhall in~\cite{lamb}. Then, if the number of
particles is kept fixed, the volume occupied by all nucleons in
all pasta-like structures will be approximately the same
irrespective of the value of the number density.

\begin{figure}[h!] %figure 11
 \centering
 \includegraphics[width=3.45in]{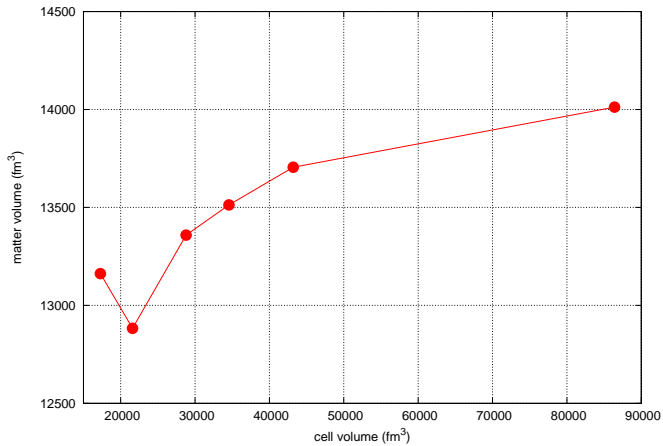}
 % sup_vs_u_cubo.png: 0x0 pixel, 300dpi, 0.00x0.00 cm, bb=0 0 640 480
 \caption{Volume occupied by matter vs. cell's volume for the Medium
 Pandharipande potential for the non-homogeneous densities.}
 \label{fig_vol_frac}
\end{figure}

Such effect can be corroborated by estimating the total volume
occupied by the nucleons (though the digitalization process
described in \cite{dor12A}) and comparing it to the total volume
of the cells, which increases with decreasing density.  Figure
\ref{fig_vol_frac} shows that the total volume occupied by
nucleons in the non-homogeneous configurations varies by less than
$10\%$ while the cell's volume increases almost ten times. In
other words, as the pasta-like structures change from drops to
rods, slabs, etc. the total volume occupied by nucleons remains
practically constant. This indicates that the structures do not to
minimize their bulk (volume) energy when they change shapes, and
points to the surface energy as the critical factor in determining
which specific pasta-like structure is the most appropriate for a
given density.

To quantify the variation of the surface area of the structures in
terms of the cell size we perform a simple geometric exercise. We
calculate the surface area of simple shapes (spherical drop,
cylindrical rod, slab and their bubble counterparts) and plot it
as a function of the fraction of the cubic cell's volume they
occupy, $u$.

Each pasta-like shape can be characterized by its volume and
surface area, which in turn can be expressed in terms of the cubic
cell's side $L$ and a characteristic length $a$, namely the radius
of a spherical drop, cylindrical rod, the width of a lasagna slab,
etc.

It must be remembered that for finite systems with periodic
boundary conditions the simulation cell imposes some constrains.
Because of the periodic boundary conditions, both slabs and
cylindrical rods (or bubbles) have some faces attached to the
surfaces of the cubic cell in which they are inscribed. These
lateral faces produce artificial surfaces and should not be taken
into account.

In this way, the effective surface area of each shape can be
written in terms of $a$ and the cubic cell's length $L$ and its
variation with the volume fraction $u$ can be studied. The surface
area of a single shape per cell as a function of $L$ and $u$ is
found to be
\begin{align*}
 S_{sphere} &= 4\pi\left( \frac{3}{4\pi} \right)^{\frac{2}{3}} \times
u^{\frac{2}{3}}\times L^2 \\
 S_{rod} &= 2\left(\pi \right)^{\frac{1}{3}} \times
u^{\frac{1}{3}}\times L^2 \\
 S_{slab} &= 2 \times L^2 \ .
\end{align*}
The bubble counterparts have similar expressions with $u$
replaced by $(1-u)$.

\begin{figure}[h!] %figure 12
 \centering
 \includegraphics[width=3.45in]{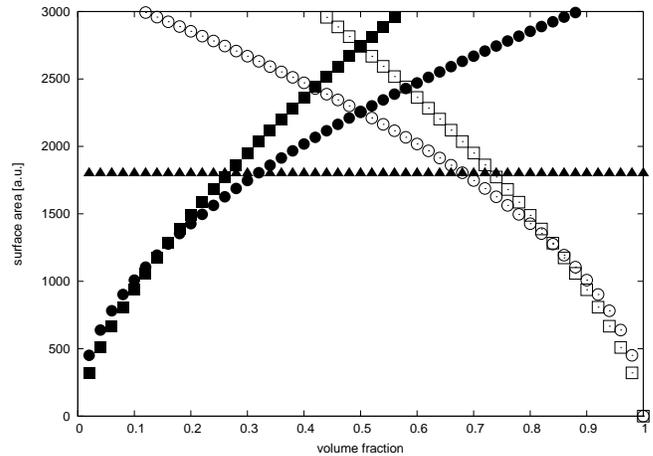}
 % sup_vs_u_cubo.png: 0x0 pixel, 300dpi, 0.00x0.00 cm, bb=0 0 640 480
 \caption{Surface area of various simple shapes as a function
 of volume fraction for a cell of $L=30$. The shapes are:
 spherical drop (full squares), cylindrical rod (full  circles), slab (full triangles),
 cylindrical bubble (empty circles) and spherical bubble (empty squares)}
 \label{fig_sup_min}
\end{figure}

Since the surface areas of all shapes studied scale as $L^2$,
there will not be a specific shape that will have a minimum
surface area for a given cell length $L$. Consequently, the shape
structures that will be dominant at a given density will be
selected entirely by their volume fraction $u$.
Figure~\ref{fig_sup_min} shows these surfaces as a function of the
volume fraction.

In order of increasing volume fraction, the preferred shapes
(minimum surfaces) go from spherical drop, to cylindrical rod,
to slab, to cylindrical bubble and finally to spherical bubble;
basically the same ordering found in almost every study of nuclear
pasta.

\begin{figure}  %figure 13
\begin{center}
\includegraphics[width=3.45in]{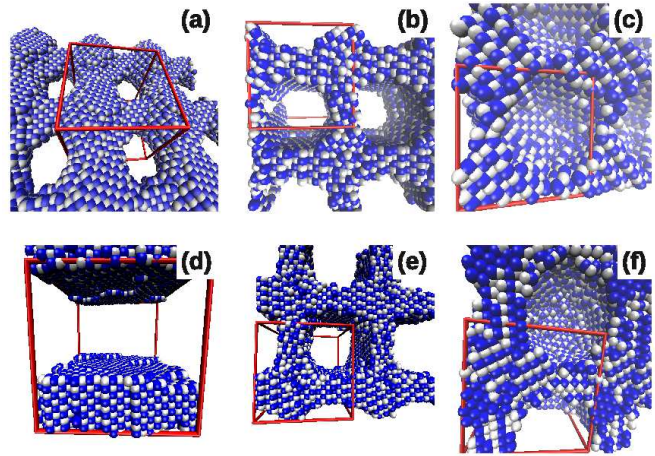}
\end{center}
\caption{Comparison of structures for the Pandharipande Medium
potential with $A=1728$ (top row) and $A=4096$ (bottom row) for
densities $\rho=0.05fm^{-3}$ (panels a and A), $\rho=0.08fm^{-3}$
(panels b and B) and $\rho=0.1fm^{-3}$ (panels c and C)}
\label{fig_varios_tamagnos}
\end{figure}

This schematic result should be exact for large enough systems,
i.e. for cell sizes much larger than the range of the interaction
potential where interfacial and curvature effects can be
neglected. Of course, in order to determine the most stable
configuration a figure such as~\ref{fig_sup_min} should be
constructed using every possible surface. The question then is,
how small is large enough? To address this question we now turn to
a study of the punctured slab.

Figure~\ref{fig_varios_tamagnos}, shows configurations found with
Pandharipande's medium potential at $T=0.1 \ MeV$ and $\rho=0.05
fm^{-3}$ and $0.08 fm^{-3}$ for three system sizes: $A=1728,\
4096$ and $13824$. It is clear that only one structure per cell is
found, independently of system size. For density $\rho=0.08\
fm^{-3}$, the same structure (cylindrical bubble) is observed for
the three sizes, but for density $\rho=0.05\ fm^{-3}$, however,
the smallest system forms a single punctured slab (PSlab) with a
single hole, whereas the larger systems form a single regular
slab. Such structure appears to be the minimum energy
configuration between a ``rod'' and a ``slab'' over a very small
range of densities, becoming a slab by adiabatically increasing
the density in a larger cell. A suggestively similar situation was
observed by Williams and Koonin~\cite{koonin} who reported a
``slab with regular holes''.

To compare the surface area of this structure to that of the slab
it is necessary to use the width of the slab ($a$), and the radius
of the holes $b$ as independent parameters. In terms of the volume
fraction $u$ it is
\begin{align*}
 b      &= L\times\sqrt{1-\frac{L\times u}{a}} \\
 S_{Pslab} &= 2L^2\times \frac{L\times u}{a} + 2\pi b\times a
\end{align*}

\begin{figure}  %figure 14
\begin{center}
\includegraphics[width=3.45in]{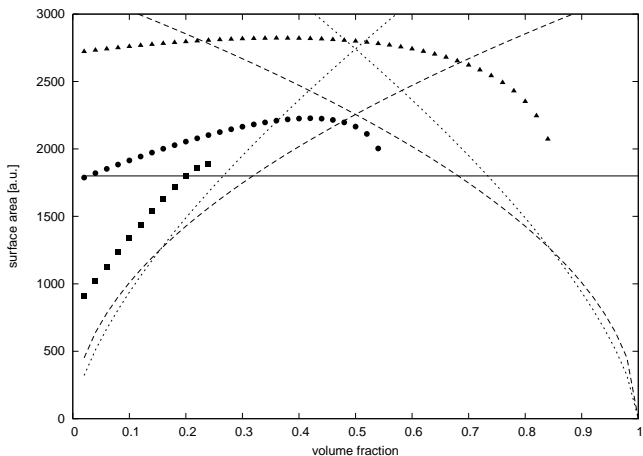}
\end{center}
\caption{Surface area of punctured slabs of width $w=0.25L$
(Squares), $w=0.55L$ (circles) and $w=0.85L$ (triangles) for
$L=30$. Surface area of other shapes are included for comparison:
sphere and spherical bubble (dotted line), rod and cylindrical
tube (dashed line) and slab (full line)}
\label{fig_sup_lasagna_agujereada}
\end{figure}

As shown in Figure~\ref{fig_sup_lasagna_agujereada}, the punctured
slab is never the minimum surface shape among those structures
considered. Furthermore, the total surface area of the punctured
slab does not scale as $L^2$ as all the others. The leading term
actually scales as $L^3/a$ which is always larger than $L^2$, so
for larger values of $L$ this shape will always have a larger
surface than the other shapes considered; yet our simulations
yielded one. This happens because small cells yield thin slabs in
which the particles at one surface could interact with those at
the opposite surface; such effect would occur in slab with
thicknesses comparable to twice the range of the interaction
potential.

In such case higher order surface terms become relevant. Indeed,
the punctured slab with $A=1728$ from Fig.
\ref{fig_varios_tamagnos} has a maximum width of $\sim 20 \ fm$
and a minimum of $\sim 10 \ fm$, while the potential has a range
of $r_c = 5.4\ fm$. The regular slab with $A=4096$ has a constant
width of $\sim 16 fm$. For $A=1728$ and $\rho=0.05\ fm^{-3}$, the
cubic cell's length is $32.6 fm$, while the range of the potential
is $r_c=5.4\ fm$. That is, the cell is only $6$ times larger than
the range of the potential, hence interfacial or curvature effects
might be of the same order than surface effects.

\begin{figure}  %figure 15
\begin{center}
\includegraphics[width=3.45in]{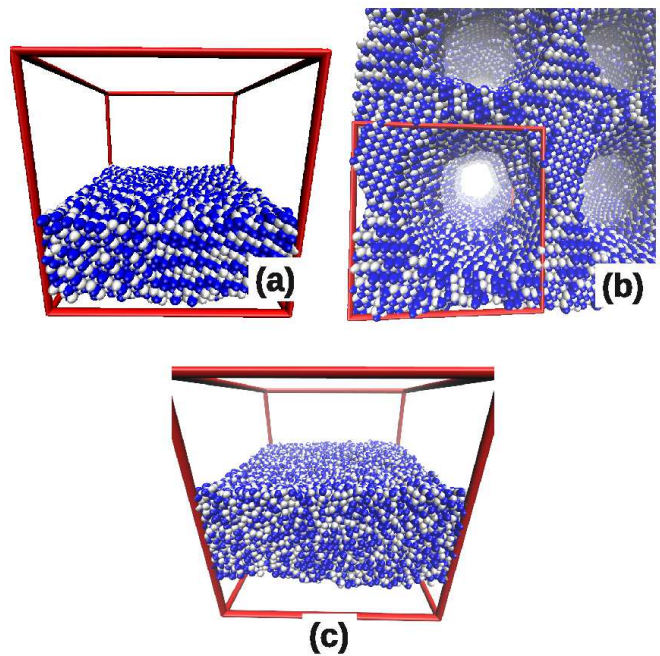}
\end{center}
\caption{Some results with larger systems. Panel (a) shows a $T=0.001 MeV$ 
configuration for $A=13824$ at $\rho=0.05fm^{-3}$, (b) $T=0.001 MeV$, 
$A=13824$ and $\rho=0.08 fm^{-3}$. Panel (c) shows a configuration for 
$A=46656$ at $\rho=0.05fm^{-3}$ and $T=0.8 MeV$}
\label{fig15}
\end{figure}

This same interplay of geometrical parameters might be responsible
for the more exotic structures shown in Figures~\ref{fig2},
\ref{figT1}, \ref{horostruc} and \ref{stiffstruc}. At any rate,
$A=1728$ is not large enough to shake off finite size effects, due
only to the nuclear interaction. The scenario is expected to be
even worse with a long range interaction such as Coulomb.

For the sake of argument, we show in fig.\ref{fig15} results obtained 
with $A=13824$ and $A=46656$ particles. At density $\rho=0.05 fm^{-3}$ 
we found slab-like structures for all system sizes equal or larger 
than $A=4096$.
For density $\rho=0.08 fm^{-3}$, with $A=13824$ almost 
cylindrical holes are observed, with some modulation. That modulation 
might be due to fast cooling.
Incidentally, the slab for $A=46656$ is stable at higher temperatures 
than for smaller systems. Compare, for example, fig.~\ref{fig2} and 
fig.~\ref{figT1}: For $A=1728$ the $T\sim0MeV$ solution is a slab, but it 
becomes distorted at higher temperatures.

A rule of thumb to distinguish pseudo-pasta --i.e. that due to
periodic boundary conditions-- and ``true'' pasta (that arising
form a balance between nuclear and Coulomb interactions) is that
the scale of the pseudo-pasta structures is set exclusively by the
size of the cell. This is evidenced by the fact that, without
Coulomb interaction, there is systematically a single structure
per cell. Any model that includes Coulomb interaction and aims to
produce ``true'' pasta should, at least, be able to produce more
than one structure per simulation cell.

\section{Concluding remarks}\label{concluding}

Nuclear matter was studied at subsaturation densities and low
temperatures using molecular dynamics using three different pairs
of two-body potentials corresponding to different values of the
compressibility. The average binding energy per nucleon as a
function of the density indicated that, around saturation density,
crystalline arrangements of the type simple cubic ($B1$), body
centered cubic ($B2$) and diamond ($B3$) crystal lattices were
found in simulations using the potentials with compressibility of
$283 \ MeV$, $418 \ MeV$ and $494 \ MeV$. These lattice structures
were observed in the temperature range of $0.001 \ MeV \lesssim T
\lesssim 1.0 \ MeV$. A departure from the crystalline structure
into non-homogeneous structures occurred very consistently
whenever the density decreased below $\rho\approx 0.13 fm^{-3}$;
this effect was observed at all temperatures studied and for the
three potentials.

The structures formed correspond to the usual pasta-like
structures that have been observed in previous studies. The fact
that these pasta shapes were formed in systems without the Coulomb
interaction (i.e. not through nuclear-Coulomb frustration)
prompted a study of the role of the size cell and periodic
boundary conditions on the formation of the structures. After
demonstrating that pasta-like structures can be obtained in a
one-component Lennard-Jones system and that the volume energy
played a minor role in the selection of the shapes, surface
effects were investigated in turn. A simple geometric calculation
showed that the unavoidable use of periodic boundary conditions
combined with the finite size of the system favors the creation of
pasta-like structures. These surface effects are not exclusive to
this study nor to a particular interaction potential, in our case
we found them to be present in systems as large as $13824$
particles. We believe that these effects might still play a
significant role in the formation of pasta even with a Coulomb
interaction; these and other related issues will be tackled in an
upcoming contribution~\cite{pedro}.

\begin{acknowledgments}
C.O.D. thanks partial financial support by CONICET through Grant PIP0871,
P.G.M. and J.I.N. are supported CONICET scholarships, and J.A.L. by NSF-PHY
Grant 1066031. C.O.D. and
J.A.L. thank the hospitality of the Nuclear Theory Group of the Nuclear
Science Division of The Lawrence Berkeley National Laboratory where this
work was initiated.
\end{acknowledgments}

\end{document}